\documentclass[preprint,12pt]{elsarticle}

\usepackage{amssymb}
\usepackage{amsmath}

\journal{Planetary and Space Science}

\begin{document}

\begin{frontmatter}

\title{Preparing for the 2061 return of Halley’s comet \\
A rendezvous mission with an innovative imaging system} 

\author[label1]{Cesare Barbieri} 
\affiliation[label1]{organization={University of Padova and INAF associated},
            addressline={vicolo Osservatorio 3}, 
            city={Padua},
            postcode={35100}, 
            country={Italy}}
            
\author[label2]{Alessandro Beolchi} 
\affiliation[label2]{organization={Department of Aerospace Engineering, Khalifa University of Science and Technology},
            addressline={P.O. Box 127788}, 
            city={Abu Dhabi},
            country={United Arab Emirates}}

\author[label3]{Ivano Bertini} 
\affiliation[label3]{organization={Department of Science and Technology, University of Naples "Parthenope"},
            addressline={Centro Direzionale Site isola C4}, 
            city={Naples},
            postcode={80143}, 
            country={Italy}}
            
\author[label4]{Vania Da Deppo} 
\affiliation[label4]{organization={CNR-IFN Padova and INAF associated},
            addressline={via Trasea 7}, 
            city={Padua},
            postcode={35231}, 
            country={Italy}}
            
\author[label2]{Elena Fantino} 

\author[label2]{Roberto Flores} 

\author[label5]{Claudio Pernechele} 
\affiliation[label5]{organization={INAF Astronomical Observatory of Padua},
            addressline={vicolo Osservatorio 3}, 
            city={Padua},
            postcode={35100}, 
            country={Italy}}
            
\author[label2]{Chiara Pozzi} 

\begin{abstract}
The return of Comet 1P/Halley will promote a worldwide interest for ground and space observations of a celestial body of outstanding scientific and cultural interest. In addition to remote observations, space will open the possibility of in situ observations, similarly to what was done during the passage of 1986. In this paper, we first discuss the scientific motivations for a rendezvous mission, capable to overcome the limitations of the flyby missions that took place at that time.
In the second part, we describe an example of a rendezvous trajectory that can be carried out with existing power and propulsion technologies, i.e., with radioisotope thermoelectric generators and a Hall effect thruster. Furthermore, the transfer is made possible by the gravitational assistance of a giant planet. The resulting mission concept, nicknamed HCREM (Halley Comet REndezvous Mission), selected from a number of cases treated in a previous paper of ours \citep{Beolchi2024}, will be capable to reach the comet beyond the distance of Saturn, when the sublimation of super-volatile species (e.g. CO and CO\textsubscript{2}) will be ongoing, and well before the onset of the sublimation of water (expected to occur around 4 AU, namely at larger heliocentric distance than Mars). 
Following a direct transfer from Earth, a gravity assist with Jupiter inserts the spacecraft into the cometary orbital plane with retrograde motion. Electric propulsion modifies the trajectory so that the spacecraft reaches the comet with zero relative velocity. After the rendezvous, the S/C will accompany the comet for several years before, around and after the perihelion, which will happen in July 2061, until the outbound crossing of the ecliptic and possibly even later, at the end of that year. Given the large heliocentric distances reached by the spacecraft, our concept mission does not foresee the implementation of solar panels. In this way, some shortcomings deriving from the adoption of this technology onboard the Rosetta mission to comet 67P are avoided and operations can occur even inside the dense dust coma at short distance from the nucleus
In the third part of the paper, an innovative imaging system is proposed, with a very large field of view of about 100°, capable to record on the same frame details on the surface and the surrounding space, in order to  follow for several degrees the trajectories outside the borders of the nucleus of chunks and clouds ejected by pits or fractures, all phenomena crucial to the understanding of the cometary activity.
In the Conclusions, we stress that a concerted effort is needed in the current decade to plan and approve a rendezvous mission to 1P. Indeed, the scenario here described requires launching before 2040, less than 15 years from now. Later launches with existing rockets imply a severe loss of scientific knowledge, because the spacecraft will not be able to reach the comet before the onset of water sublimation.

\end{abstract}


\end{frontmatter}

\section{Introduction}
\label{intro}

After its passage through the aphelion in December 2023, comet Halley, last seen in 2003 \citep{Hainaut2004}, will return near the Earth in July 2061. According to the prediction contained in https://www.spacereference.org/comet/1p-halley?, the next perihelion will occur on 28 July 2061. Based on such orbit, the expected dates for its next return journey are the following:
\begin{itemize}
\item Passage at the distance of Saturn: December 2058. 
\item Passage at the distance of Jupiter: June 2060 (Jupiter MOID = 0.7828 AU**).
\item Minimum distance to Jupiter: 9 Sept 2060 at 0.9835 AU.
\item Passage at the distance of Mars: May 2061. 
\item Ingress on the ecliptic: May 2061.
\item Passage at the distance of the Earth: June 19, 2061, Earth MOID = 0.0744805 AU**.
\item Passage at the distance of Venus: July 9, 2061.
\item Perihelion: July 28, 2061, at 0.5928 AU distance from the Sun, $\sim$ $88.7 \times 10^6$ km.
\item Minimum distance to Earth: 29 July 2061 at 0.477 AU, $\sim 71.4 \times 10^6$ km, with the very high relative velocity of  83.4 km/s.
\item Minimum distance to Venus: 20 Aug 2061 at 0.05435 AU, $\sim 88.7 \times 10^6$ km*.
\item Egress from the ecliptic: September 2061.
\item Outbound passage at the Mars distance: October 2061.
\end{itemize}

*data retrieved from (solution date 2024 April 15):
\begin{verbatim}
https://ssd.jpl.nasa.gov/tools/sbdb_lookup.html#/?sstr=1P&view=OPC
\end{verbatim}

**MOID: The minimum orbital intersection distance (MOID) at a given epoch between two objects orbiting a common primary is the distance between the closest points of their osculating orbits.

Around the perihelion of 2061, the comet will be better positioned for observation from terrestrial observers than during the 1985-1986 apparition, as it will be on the same side of the Sun as the Earth. Notice that the closest approach to Earth will be one day after perihelion. 

Although the crucial phases of its ingress in the inner Solar System are still more than 30 years in the future, we started to examine the feasibility of a space mission using present-day rockets and technologies, with the aim to delineate a scenario where the comet can be reached well before the onset of water vapour release, which happens around 4 AU \citep{Fulle2022}, namely before the arrival near the orbit of Mars. Indeed, we shall show in Chapter \ref{ch3} that departing toward the end of the next decade (e.g. 2037), the comet can be reached even before the orbit of Jupiter or even Saturn. Such mission should be capable to solve several questions related to the comet itself and to cometary science in general.

The paper is articulated as follows:
\begin{itemize}
\item Chapter \ref{ch2} motivates on scientific grounds the need for a rendezvous and not a flyby mission. 
\item Chapter \ref{ch3} illustrates an example of the trajectories investigated for such mission.
\item Chapter \ref{ch4} discusses an innovative imaging system capable to overcome the limitations of previous systems.
\item In the Conclusions, we stress that launches later than 2040 with existing rockets imply that the spacecraft (hereafter S/C) will not be able to reach the comet before the onset of water sublimation, with a severe loss of scientific knowledge.
\end{itemize}

Many of our considerations rely on the experience gained with previous cometary missions, in particular with the GIOTTO flyby in 1985/86 to comet 1P/Halley and the ROSETTA rendezvous with comet Churyumov-Gerasimenko (67P/C-G) in the period 2004-2016.  In addition, the forthcoming ESA Comet Interceptor mission has given us the possibility to study the innovative imaging system described in Chapter \ref{ch4}.

\section{Scientific interest for a space mission covering the passage of 2061}
\label{ch2}

We discuss here a number of scientific motivations to observe the comet in its next return, from ground and space remote spacecraft and in situ missions. We enumerate firstly some motivations related to the comet itself; subsequently, motivations of a more general character will be given.

In March 1986, a truly “Grande Armada” of spacecraft were directed to comet Halley. Two spacecraft from Japan (Suisei, Sakigake), two from URSS (Vega 1, Vega2) and one from ESA (GIOTTO) had the comet as their specific target. NASA diverted an existing spacecraft (ICE) to take data from a distance.  

The coordination among the Agencies was insured by the Inter Agency Consultative Group (see https://sci.esa.int/web/home/-/29231-inter-agency-consultative-group-returns-to-its-birthplace-for-twentieth-anniversary-celebrations).

Although there was no specific NASA mission, USA researchers had a great role in preparing for the return of 1P, for instance in the coordination of ground and space observations insured by the International Halley Watch (see https://www.jpl.nasa.gov/news/international-halley-watch/). In particular, we acknowledge their presence in the hardware and software of the Halley Multicolour Camera (HMC) on board the GIOTTO mission \citep{Keller1986}. The GIOTTO HMC images had a resolution of 45 m/px at the minimum distance from the comet (about 600 km). Also the imaging systems of the two VEGAs took images of the nucleus, but from a larger distance and of lesser quality due to several problems (see for instance \cite{Stooke1991}). For reprocessed HMC and Vega2 images see: https://apod.nasa.gov/apod/ap100104.html and https://www.planetary.org/space-images/the-nucleus-of-halley-s-comet, respectively. The images proved conclusively that comets have indeed the nucleus theorized by several astronomers (in particular at the time by \cite{Whipple1987}) and allowed the measurement of several properties of it, such as dimensions ($\sim 8 \times 8 \times 16$ km), low albedo (0.04), small fraction of active area, content of refractories, organic material and ices, high temperature at the date of the flyby ($\geq$ 360 K), high porosity and low tensile strength. Moreover, the high abundance of the super-volatile gases CO and CO\textsubscript{2} pointed out that the formation of the nucleus took place in an ambient of very low temperature, below 30 K \citep{Keller1990}.  

However, several important characteristics of the nucleus remained poorly known or entirely unknown, such as:
\begin{itemize}
\item The shape of the entire body and the geomorphology of the surface.
\item The localization and dimension of active regions.
\item The determination of the complex rotation state.
\item The amount of mass loss. 
\item The internal structure.
\item The accurate mass spectrometry of the gaseous and solid components.
\end{itemize}

Regarding the morphology, the cited Vega2 image suggests a bilobate structure, reminding of that of 67P taken by the OSIRIS/Rosetta imaging system \citep{Sierks2015}, and indeed the majority of cometary nuclei observed from space and ground-based radio telescopes showed a bilobate structure (see for instance Fig. 1 in \cite{Keller2020} and the nice collection of space and radar images at 
\begin{verbatim}
https://planetary.s3.amazonaws.com/web/assets/pictures/20181218_Comets25mpixi4.png
\end{verbatim}

The hypothesis that Halley’s nucleus is a contact binary can be substantiated only by future high resolution images covering the entire body and by the determination of the internal structure by navigation and possibly penetrating radars such as CONSERT/Rosetta. Images with high spatial resolution on one side and panoramic view on the other, with high temporal cadence, seem indispensable to the better definition of the cometary properties. This is why we concentrate in Chapter \ref{ch4} to describe an innovative imaging system.

Now, we point out some motivations of a general character to observe Halley’s in 2061.
Rosetta provided an extraordinary amount of data on comet 67P, some of them specific to that comet, others applicable to the entire cometary science and even to astrobiology, because all comets investigated with sufficient detail seem to carry material of interstellar/pre-solar origin as well as solar processed compounds. In other words, a comparison with other comets visited by spacecraft to date seems to indicate a basic similarity of all comets, masked by the different age and evolutionary history. Indeed, the major constituent is always refractory material formed at very low temperatures in interstellar nebulae. The D/H value measured by ROSINA/Rosetta \citep{Altwegg2015} taken at bare face seem to indicate that comets have contributed to the inert gases of the pristine Earth’s atmosphere but not to the majority of its watery content. However, a recent study reopens the question, because it is shown that dust increases the locally measured D/H ratio \citep{Mandt2024}.

And yet, important questions remain in need of additional investigation, such as a detailed composition of the refractories and organics, measurements of a possible difference between the refractory to ice ratio in the nuclei and the dust/gas ratio in the coma and between the D/H ratio in the pristine nucleus and in the coma, the relative importance of CO+CO\textsubscript{2} sublimation versus H\textsubscript{2}O sublimation in controlling activity and erosion, and above all, the physics of the release of internal energy responsible for the activity. 

Regarding this latter point, namely the release of internal energy, comet Halley had a dramatic increase in brightness on Feb. 12, 1991. A that time, the comet was already at 14.3 AU from the Sun (in between the orbits of Saturn and Uranus) and about 18° below the ecliptic plane. The outburst was discovered by O. Hainault and A. Smette on plates taken with the 1.5 m Danish Telescope at the ESO-La Silla Observatory (see the ESO Press Release, https://www.eso.org/public/news/eso9103/). The 300-time increase of brightness was followed by other observers until the end of March. Many explanations were put forward for such unexpected phenomenon, such as an impact with a solar shock wave \citep{Intriligator1991}, sublimation of CO and CO\textsubscript{2} \citep{Meech1991,Sekanina1992}, impact with an outer belt asteroid \citep{Hughes1991}, while \cite{Weissman1991}, after reviewing other possible explanations, pointed out the possible role of amorphous ice in transforming to the crystalline phase.

At any rate, when the comet was last seen in 2003, at heliocentric distance $r = 28.1$ AU, its brightness had returned to the expected, nucleus-only, value, $R=28.22 \pm 0.13$ \citep{Hainaut2004}. The coordinates of the comet differ from 0.3 to 1.4 arcsec from the JPL Horizon ephemeris, but these measurements are not taken into account in the current orbit (denoted as IAU74) quoted in the Introduction \citep{Giorgini2024}.

Although external sources of energy such as solar wind and impact with asteroids are a possibility, the source of internal energy and its manifestation as activity periods (and for that matter of inactivity periods) remain the biggest puzzle. We still do not know how the interplay of dust and gas near the surface takes place, which role water and super volatiles play for the activity, or how outbursts are triggered. In addition to explosions of the super-volatile gases, the already mentioned process of phase changes between amorphous and crystalline ice has been variously discussed, but with no clear conclusions. Indeed, amorphous ice is expected to be the phase of ice formed at temperatures and pressures typical of the interstellar medium. Its presence and importance were discussed already by \cite{Smoluchowski1981}, but it is very difficult to detect it, as discussed for instance by \cite{Prialnik2022}.

Observations of comets give evidence that several bodies display activity well beyond the water ice sublimation boundary (e.g. \cite{Womack2017} and references therein). This is generally explained as the consequence of the sublimation of supervolatile ices (mainly CO and CO\textsubscript{2}) at large heliocentric distances. This is indeed a fascinating topic in cometary science since it is still not known when nuclei activity starts since all available data are affected by an observational bias due to the sensitivity of past and present-day instruments and detectors.

If we take into account sublimation temperatures of cometary species \citep{Womack2017} and models of the blackbody temperature of cometary nuclei versus heliocentric distance \citep{Ferrin2013}, it can be easily derived that CO\textsubscript{2} starts subliming around 20 AU from the Sun, while the supervolatile CO may drive cometary activity well beyond 100 AU.

To understand those basic questions about internal energy and activity, data with high spatial resolution are needed to determine the microstructure and small-scale strength, the thermal inertia and heat conduction coefficient of the surface layer, the distribution of ice and dust below the surface. Answers to those questions require in situ observations, and specifically a rendezvous and not a flyby mission, as motivated in detail in Paragraph \ref{rendezvous}. The composition of the scientific payload of such a mission is surely premature. The presence of a lander (and possibly more than one) is highly desirable in addition to the scientific complement on the orbiter. A sample return mission would be of utmost importance, as pointed out for instance by \cite{Thomas2019} in the frame of a general discussion of future space missions. We do not discuss these highly complex configurations. In Paragraph \ref{requirements} we limit ourselves to identify few top-level characteristics of the S/C and describe an imaging system part of payload on the orbiter.

\subsection{Why a rendezvous and not a flyby mission to Comet Halley}
\label{rendezvous}

Our decision to study a rendezvous and not a flyby mission to comet 1P is motivated by the previous considerations and, in particular, by the results obtained by ESA/Rosetta, that could follow the evolution of cometary surface and activity for over two years, from its  insertion in orbit around the nucleus that took place near the orbit of Jupiter in the inbound trajectory, through perihelion and beyond the orbit of Mars on the outbound leg. These results could not have been achieved by flyby missions. Here are some of these results obtained by the imaging system OSIRIS, composed by a Narrow Angle and a Wide Angle cameras: 
\begin{itemize}
\item A confirmation of the importance of the orbital parameters of the comet, such as position and changes of the rotation axis with respect to the orbital plane and of the diurnal rotation measured before and after the perihelion \citep{Mottola2014,Godard2015,Kramer2019}.
\item Seasons could be precisely associated with the nucleus evolution, in particular to the transfer of mass between the two hemispheres \citep{Keller2017}.
\item The accurate determination of the geomorphology of the entire surface \citep{Thomas2018} and the onion-like shell structure suggesting a bilobate shape \citep{Penasa2017}.
\item The calculation of the mathematical model of the overall shape to better than few meters \citep{Preusker2015,Preusker2017}.
\item The time variation of features on the surface, such as boulders disaggregation and sand movements, collapse of cliff etc. (see e.g., \cite{Pajola2017}).
\item Appearance and disappearance of frost at the terminator \citep{Fornasier2016}, the release of large chunks in the centimeter to decimeter range \citep{Fulle2016}, deep pits with sublimation from their walls \citep{Weissman2015,Vincent2015}.
\item The ratio of very low tensile-strength (less than 100 Pa) to surface gravity (about 10-5 g) is not different from the terrestrial values, thus justifying the Earth-like aspects of mountains, cliffs and erosion \citep{Groussin2019}.
\item The dichotomy between the two hemispheres, the northern surface covered by regolith and the consolidated rocky southern \citep{Keller2020}.
\item The scattering function of dust, the most abundant material in comets \citep{Bertini2017,Bertini2019}.
\end{itemize}

Of course, of the greatest importance were the results obtained by the other instruments on the orbiter and on the lander, but as already stated we focused on data allowed by the long navigation around the nucleus of the imaging system OSIRIS. In addition to the already quoted paper by \cite{Keller2020}, see the review by \cite{Fulle2016}. The above selection has been done in order to suggest a novel imaging system for HCREM expounded in Chapter \ref{ch4}.

\subsection{Top-level scientific requirements of the S/C}
\label{requirements}

HCREM must reach the comet well before the onset of water sublimation (estimated to occur before or around 4 AU, e.g., \cite{Fulle2022}), namely well before the orbit of Mars. Even better, if the comet could be reached near the orbit of Saturn around 10 AU, where we expect repeated bursts of CO+CO\textsubscript{2} (the onset of production of CO\textsubscript{2} is estimated to occur around 20 AU, while CO is sublimating well beyond 100 AU). Based on the experience with previous space missions, and in particular with Rosetta, we derive some top-level characteristics of the S/C:
\begin{itemize}
\item A three-axis stabilized orbiter, capable to sustain a long duration mission before the rendezvous (20 years or so, Rosetta navigated for 12 years) and to follow the comet for about 2 years, from the orbit of Jupiter to perihelion to post-perihelion after the orbit of Mars (similar to Rosetta).
\item The capability to carry a substantial amount of scientific payload. For comparison, we recall the data of Giotto and Rosetta: Giotto had a total mass at launch of 960 kg and a mass of instruments of 58 kg. Rosetta had a payload mass at launch of 2900 kg inclusive of 1670 kg of propellant and a scientific payload of 165 kg on the orbiter plus 100 kg of the Philae lander.
\item The capability to measure small deviations from a very accurate determination of the orbit, to model the effects of the non-gravitational forces. 
\item The S/C must retain an amount of propellant ensuring the capability to navigate around the nucleus and hover above it to distances of the order of 1 nucleus radius (say less than 5 km) for many years, at a large range of heliocentric distances.
\item The capability to hover above the surface of the comet even inside a dense coma implies the lack of solar panels and the design of star trackers able to discriminate dust speckles from real stars.

Given those requirements, although it is premature to define the scientific complement of instruments (which should include a lander), we considered in our study a total weight of the payload of 200 kg. Moreover, we imposed a residual amount of propellant after the rendezvous capable to maintain the S/C in orbit around the comet for several years. We also foresee characteristics of HCREM which overcome some of the limitations of the Rosetta mission, such as the absence of solar panels and better star trackers, in order to have the orbiter operational inside the dust coma, down to say one cometary radius, namely about 5 km above the solid surface. Therefore, we considered nuclear power and thermocouples as the sole power system.
\end{itemize}

\section{An example of a Rendezvous Trajectory with Halley’s Comet}
\label{ch3}

The main challenges for such mission can be here highlighted:
\begin{itemize}
\item Small, light target on a retrograde and highly inclined orbit.
\item Power supply ensured by radioisotope thermoelectric generators (RTGs).
\item Long transfer time, say 20-25 years, before rendezvous.
\item Determination of thrust magnitude and direction at all times (outside ballistic arcs).
\item Trajectory design in a high-fidelity, ephemeris-based dynamical framework.
\item S/C to reach the comet at zero relative velocity (no dust shield).
\item Scientific payload $\sim200$ kg. 
\item Navigate the nucleus for at least 18 months around the perihelion (propulsion required, mass of comet too small).
\item Minimum distance above cometary surface: few km (less than 1 cometary radius).
\end{itemize}

Given those requirements and challenges, we studied a number of cases \citep{Beolchi2024} where the S/C leaves the Earth's vicinity along a hyperbolic path, with large but feasible values of the launch characteristic energy C\textsubscript{3} ($<200$ km\textsuperscript{2}/s\textsuperscript{2}) (see \cite{Girija2023} for a summary of these values) and is injected with low-thrust electric propulsion arcs into 1P/Halley’s orbital plane via a Jupiter or a Saturn flyby. The gravitational pull of the planet inserts the spacecraft into a heliocentric path that minimizes the velocity difference at rendezvous. After the flyby, with the aid of electric propulsion, the trajectory is reshaped so that the velocity difference at rendezvous becomes zero. The method designed to compute the propelled trajectory uses a reference solution obtained by solving a two-body Lambert problem between the position of the flyby planet and Halley (two positions and a transfer time yield one Keplerian arc with two velocity impulses, one at each endpoint) and builds the trajectory so that the transfer time of the Lambert arc is preserved, but the thruster is active for long periods of time providing continuous variations of the S/C velocity. The thrust level is fixed, and the trajectory is numerically propagated under the gravitational effect of the Sun and the acceleration of the engine. At each instant, thrust is applied in the direction that minimizes the residual total impulse required to connect the instantaneous end points of the trajectory with a Lambert arc. For further details of the method, the reader is referred to \cite{Beolchi2024}.
For the purpose of simulating a realistic mission scenario, the rendezvous mass of the spacecraft is 1000 kg. The thrust is assumed to be entirely steerable and provided by a constant-specific-impulse (1600 s) and constant-thrust engine (36 mN). Because the thruster is activated when the spacecraft is far from the Sun (at least outside of Jupiter’s orbit), RTGs are preferred to solar panels as the source of electric power to feed the engine. 

\begin{figure}[t]
\centering
\includegraphics[scale=0.1]{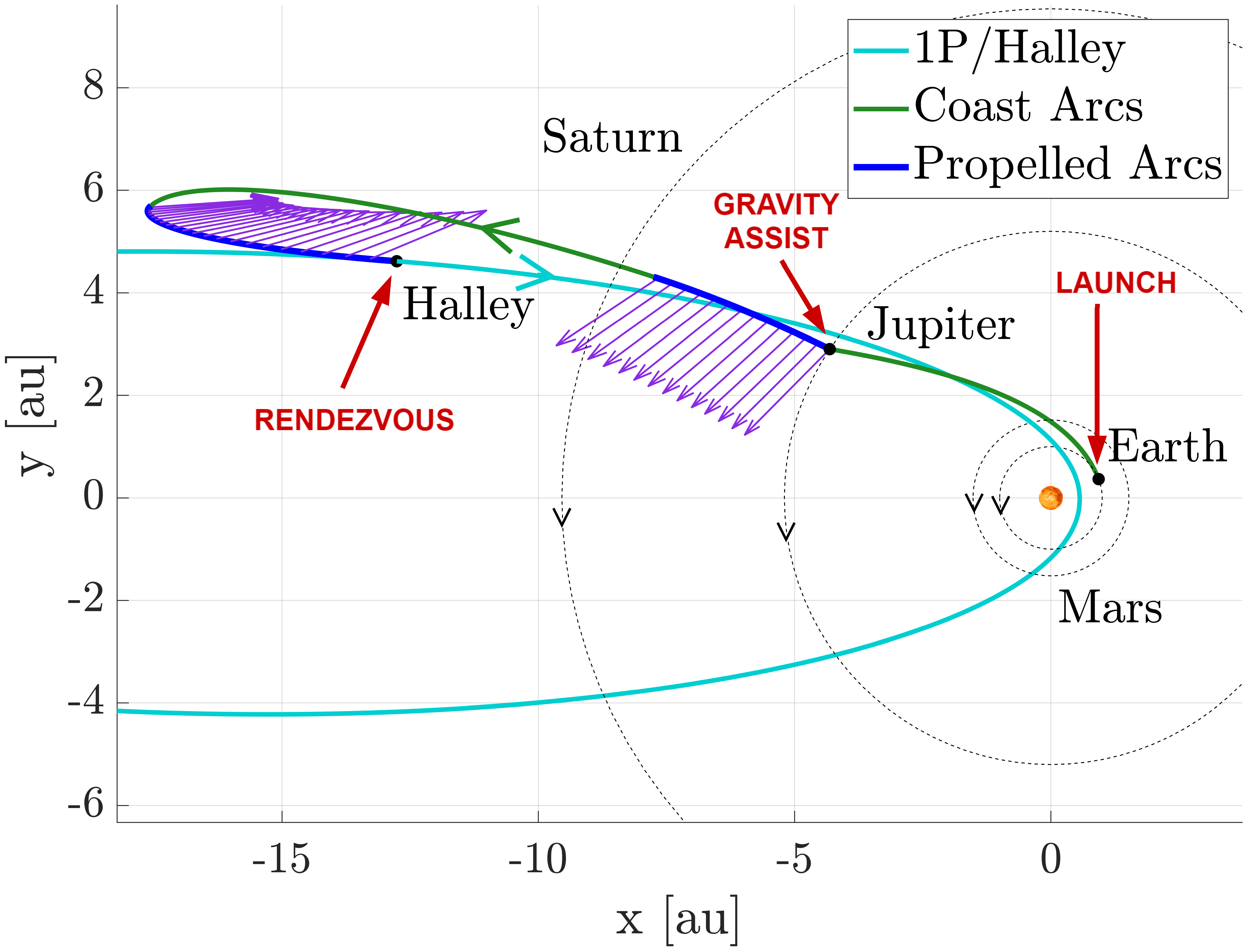}
\includegraphics[scale=0.1]{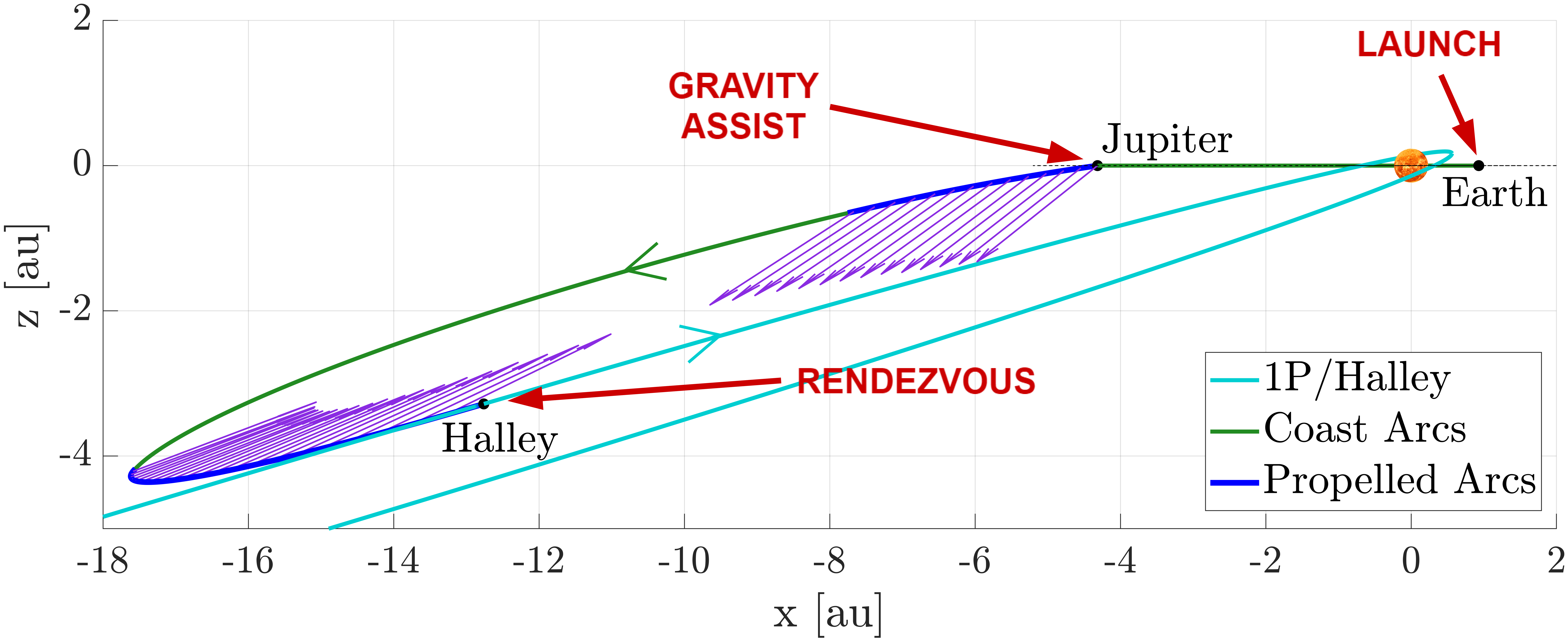}
\caption{Case Jupiter 1 trajectory (top: ecliptic view; bottom: xz plane). Arrows indicate the instantaneous direction of thrust.}\label{fig1}
\end{figure}

For sake of definiteness, among the several cases treated by Beolchi et al. (2024), we selected for the present paper the trajectory named “Jupiter 1” (Figure \ref{fig1}). This trajectory foresees departure from Earth on Aug. 18, 2036, a flyby with Jupiter on Sept. 03, 2037, and rendezvous with the comet on Oct. 02, 2056, at a heliocentric distance of $\sim 13$ AU and $\sim 6$ AU below the ecliptic plane. Therefore, HCREM will encounter the comet in the inbound leg of the trajectory at a heliocentric distance similar to that where the comet on the outbound leg had the dramatic increase of brightness discussed in Chapter \ref{ch2}.

In the example, rendezvous occurs 19 years after the Jupiter flyby. This long time can be used to the benefit of the science and the mission. Cruise science and serendipity discoveries are expected, as always happened in previous missions. First of all, on comet Halley itself, expected to become active at least at 20 AU heliocentric distance and quite possibly even earlier. HCREM will observe such activity from vantage points not accessible to ground or near-Earth space telescopes. Moreover, for most of its journey the S/C will charter the essentially unexplored Solar System below the ecliptic plane, spanning from Mars to beyond Saturn heliocentric distances. 

Such vast region of the Solar System is populated by Main Belt asteroids, Trojans, Centaurs, passing-by comets, and some of then are likely to be fruitfully imaged from closer distance and different observing geometry with respect to ground-based observations.

In the crucial phases of the mission, on 9 September 2060, HCREM will pass within 0.98 AU from Jupiter, and on 20 August 2961 within 0.0543 AU from Venus.

Additionally, this time will allow refinement of the on-board software, both operational for the S/C and of the scientific instrumentation, which could take advantage of intervening information from ground or space telescopes. Last but not least, cruise science will maintain high the interest of scientists and students that need results for papers and theses, and of the public too.

\section{An innovative imaging system}
\label{ch4}

We describe in the following an innovative all-dioptric imaging system characterized by a very wide field of view (fov), chromatic response in the range 550-800 nm, with no moving parts (no filter wheel, no shutter), fixed in the body of the spacecraft and taking advantage of the navigation to repeatedly image the entire surface. This system, meant to complement other cameras on board HCREM, will be named Very Wide Angle Camera (VWAC). For a comparison, the fov of the VWAC will be about 8 times that of the WAC/OSIRIS on board the Rosetta spacecraft. 

\begin{figure}[t]
\centering
\includegraphics[scale=0.15]{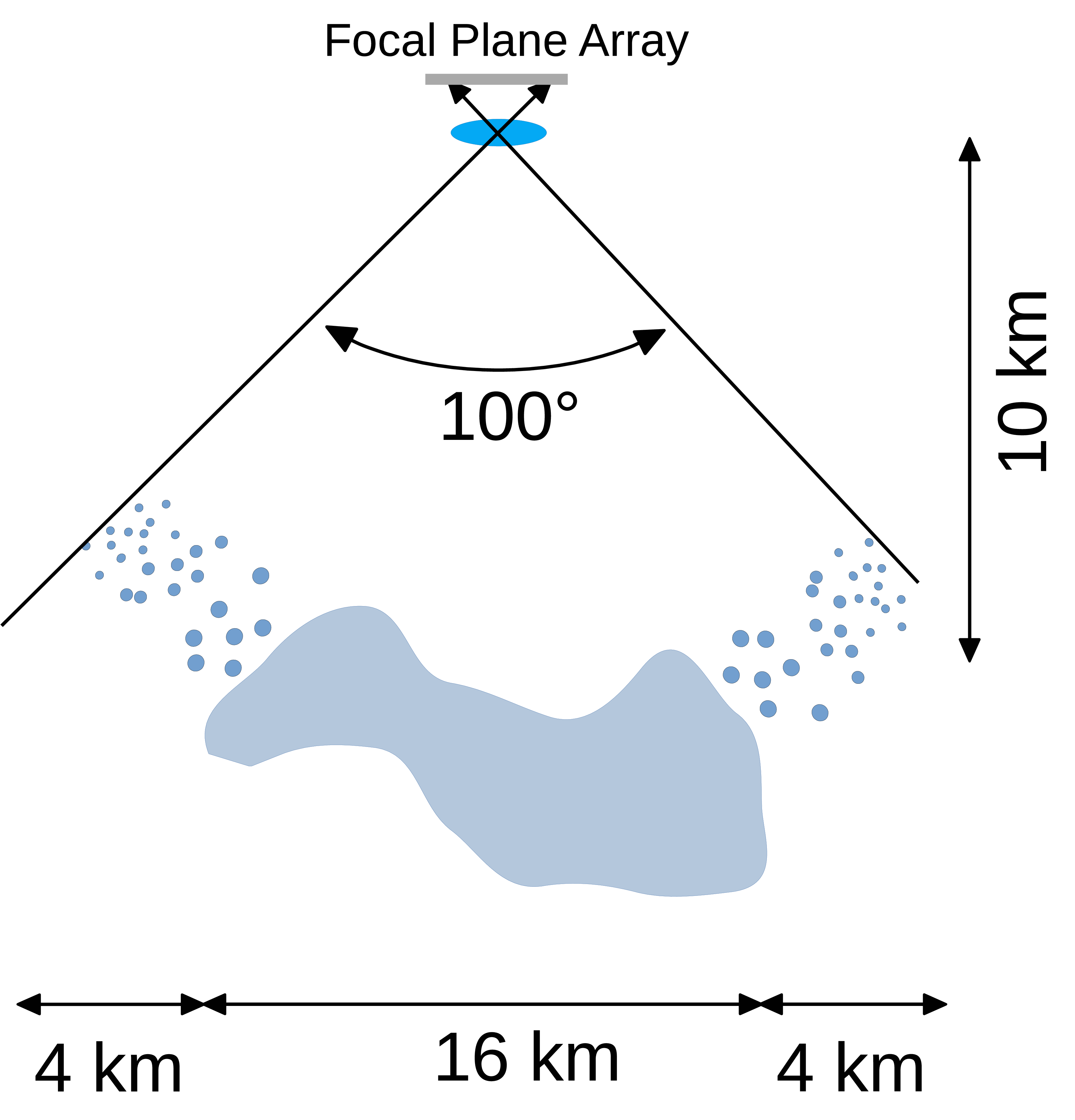}
\caption{A 100° fov lens can monitor an area with a diameter of 24 km at a distance of 10 km.}\label{fig2}
\end{figure}

To determine the fov, let us recall that the dimensions of the 1P nucleus are approximately $16 \times 8 \times8$ km. In order to image the entire nucleus from a distance down to $\sim 10$ km, an optical system with at least 92° fov is necessary. Several scientific reasons suggest to enlarge this fov, for instance to monitor outside the borders of the solid surface to detect material escaping laterally with respect to the line of sight, to study the inner coma features and their evolution, and so on. Lenses with 100° fov will be able to perform this task, monitoring up to 4 km away from the cometary body rim, as shown in Figure \ref{fig2}. 

\begin{figure}[t]
\centering
\includegraphics[scale=0.2]{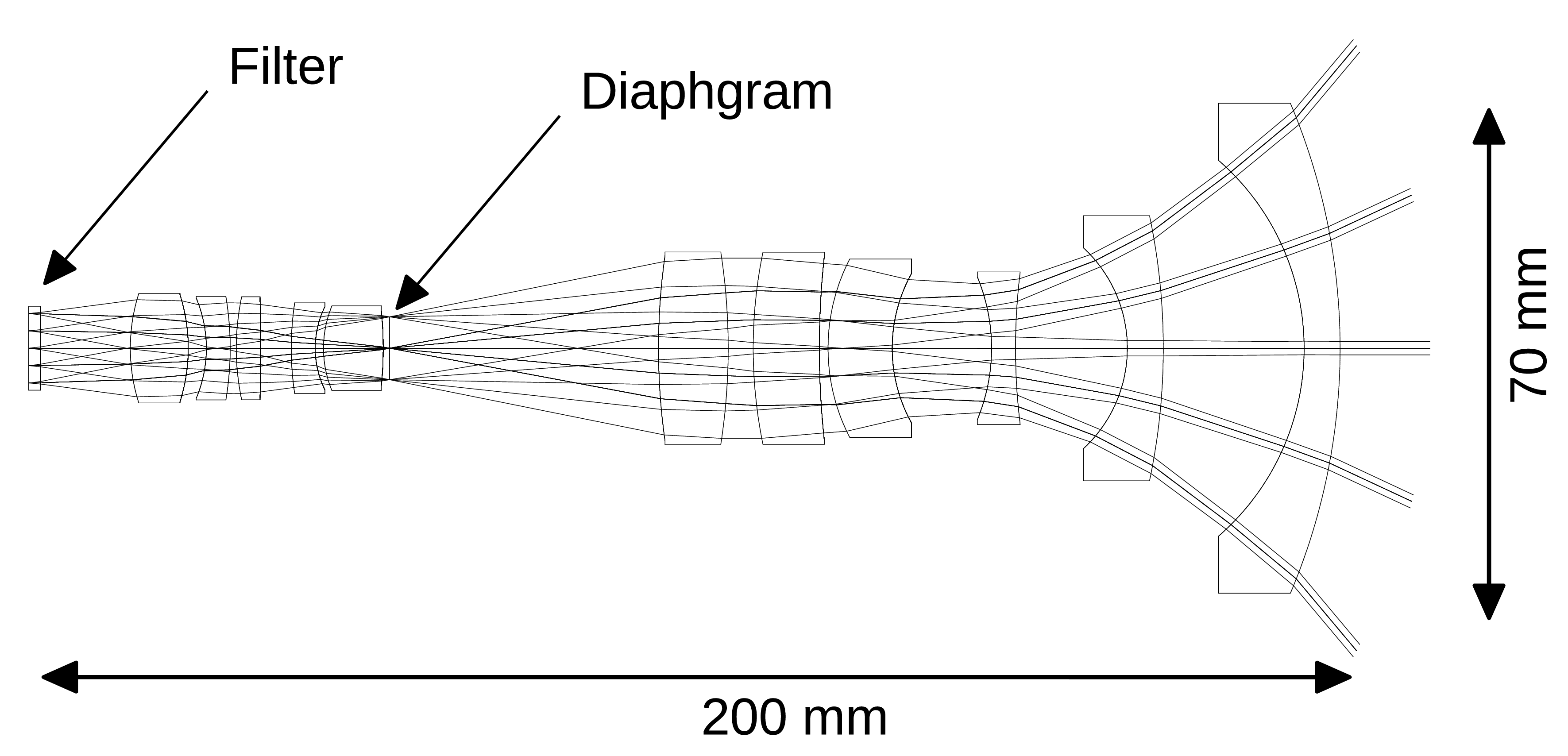}
\caption{The VWAC optical layout. A panchromatic filter is foreseen in front of the detector.}\label{fig3}
\end{figure}

In current solutions, wide angle lenses produce a distorted image, in order to fit it in a finite-sized detector, because the plate scale varies across the focal plane. The distortions though are under the control of the optical designer, depending on the wanted applications. Indeed, each distortion function has a different fov mapping function, which can be selected by the user to optimize the scientific outcome, e.g., accurate undistorted photogrammetry over the entire fov, or flux and brightness measurements maintaining linearity over a wide range.  
Recently, we overcame such limitations by designing an undistorted (F-theta) fisheye camera \citep{Pernechele2021} for the Entire Visible Sky camera (EnViss, \cite{Dadeppo2022}), meant to be mounted onboard the ESA Comet Interceptor F-class mission \citep{Jones2024}. 
The main aim of the Comet Interceptor mission is the study of a dynamically new comet, i.e. a comet that never travelled through the Solar System, or an interstellar object entering the inner Solar System. The EnVisS camera’s original design aimed to capture the entire sky in the 550-800 nm wavelength band while the spacecraft passes through the comet's tail environment.  
The F-theta design maintains distortions at the lowest possible level, thus facilitating the image analysis and photogrammetric measurements, because each sensor pixel will map exactly the same angle on the sky, no matter if the objects are on the optical axis or on the image rim. Moreover, a telecentric design assures that wide and narrow-band filters placed before the focal plane will work properly over the entire fov, although this properties in not exploited in the present application, where we only foresee a panchromatic fixed filter. 
The Very Wide Angle Camera proposed for HCREM is a downscaled version of the EnVisS designed lens (see  Figure \ref{fig3}). 

The two front lenses are made of Schott K5G20 rad-hard glass, in order to prevent transparency and colour degradation from ionizing radiation in the long mission duration. The camera focal length is 6 mm, the entrance pupil diameter is of 2 mm with a F/3 luminosity and a PSF almost diffraction limited over the entire fov.

The optical characteristics of the VWAC are summarized in Table \ref{table}. The table contains also spatial resolution values for a specific detector choice. Such off-the-shelf CMOS global shutter image sensor has a full-well charge of 13.5 ke\textsuperscript{-} and a noise of 13 e\textsuperscript{-}, allowing a nominal contrast ratio around 1:1000. Peak QE is 60\% at 550 nm. Future developments in detector technologies could permit more pixels with smaller dimensions and higher resolution.

\begin{table}[t]
\centering
\begin{tabular}{c c}
\hline 
Wavelength coverage   & 550-800 nm, 1 broad band filter (B), Panchromatic\\
Instrument FoV  & 100° \\ 
Entrance aperture (F number) & 2 mm (f/3)  \\
PSF	 & Diffraction limited \\
MTF &	>70\% at 45 lp/mm \\
Distortion	& less than 0.1\% (F-Theta) \\
Telecentricity	& less than 1° \\
Detector	 & CMOS 2k $\times$ 2k 5.5 micron px size \\
Scale factor	 & 0.05°/px \\
\hline
\end{tabular}
\caption{VWAC Optical characteristics}\label{table}
\end{table}

With the selected detector, the 100° fov permits a uniform angular sampling of approximately 0.05°/px, i.e. a plate scale of 12 m/px on the entire image, at a camera-comet distance of 10 km.  Given the small focal length, the camera will remain in good focus even to much smaller distances, with a spatial resolution improving linearly with such a decrease. 

\section{Conclusions}
\label{conclusions}

We have shown in this paper that several scientific open questions, such as the very shape of the comet, prompt for a Rosetta-type rendezvous and not a GIOTTO-type flyby space mission to Comet Halley in the next return of 2061. A specific example, named HCREM and selected from several trajectories discussed in a previous paper of ours \citep{Beolchi2024}, demonstrates that present propulsion technologies are fully capable to allow such long-duration mission, with a sizeable payload. Suggestions are made about the S/C configuration to overcome limitations that affected the ESA-Rosetta mission, like the absence on HCREM of solar panels.  Moreover, an innovative imaging system, named VWAC and characterized by the absence of distortions over a very wide field of view can be realized with off-the-shelf components. VWAC has a fov about 8 tim3s that of the OSIRIS/WAC on Rosetta, is essentially diffraction limited and has no moving part.
We stress that a concerted effort is needed in the current decade to plan and approve a rendezvous mission to 1P. Indeed, the scenario here described needs launches before 2040, less than 15 years from now. Later launches with existing rockets imply a severe loss of scientific knowledge, because the S/C will not be able to reach the comet before the onset of super-volatiles and even of water sublimation.

\section*{Acknowledgements}

The work of C. Barbieri has been partially supported by a visiting scholar contract at Khalifa University of Science and Technology under grant CIRA-2021-65/8474000413. A. Beolchi, C. Pozzi, E. Fantino and R. Flores acknowledge Khalifa University of Science and Technology's internal grant CIRA-2021-65/8474000413. A. Beolchi and E. Fantino have received support from project 8434000368 (Khalifa University of Science and Technology's 6U Cubesat mission). E. Fantino acknowledges grants ELLIPSE/8434000533 (Abu Dhabi's Technology Innovation Institute) and PID2021-123968NB-I00 (Spanish Ministry of Science and Innovation).

\bibliographystyle{elsarticle-num-names} 
\bibliography{References.bib}

\end{document}